\documentclass[aps,pra,showpacs,amssymb,longbibliography,twocolumn]{revtex4-1}
\usepackage{graphicx}
\usepackage{subfig}
\usepackage{epstopdf} 
\DeclareGraphicsExtensions{.pdf,.eps,.png,.jpg,.mps}
\usepackage{hyperref}
\usepackage{bm,amsmath}
\usepackage{dcolumn}

\newcommand{\ket}[1]{\left|#1\right\rangle}

\def\BEq{\begin{equation}}
\def\EEq{\end{equation}}
\def\BEqA{\begin{eqnarray}}
\def\EEqA{\end{eqnarray}}

\begin{document}

\title{Controlled-{\small NOT} logic gate for phase qubits based on conditional spectroscopy}

\author{Joydip Ghosh}
\email{jghosh@physast.uga.edu}
\author{Michael R. Geller}
\email{mgeller@uga.edu}

\affiliation{Department of Physics and Astronomy, University of Georgia, Athens, Georgia 30602}

\date{\today}

\begin{abstract}

A controlled-NOT logic gate based on conditional spectroscopy has been demonstrated recently for a pair of superconducting flux qubits [Plantenberg {\it et al}., Nature 447, 836 (2007)]. Here we study the fidelity of this type of
gate applied to a phase qubit coupled to a resonator (or a pair of capacitively coupled phase qubits). Our results show 
that an intrinsic fidelity of more than 99\% is achievable in 45ns.
\end{abstract}


\maketitle

\section{Introduction}
\label{sec:INTRO}

Approaches to the development of a large-scale quantum computer face numerous practical challenges \cite{ncbook,DiVincenzo2000fp,You2005pt}. One such challenge, the construction of a robust controlled-NOT (CNOT) logic gate, has  been explored from several perspectives \cite{StrauchPRL03,ZhangPRA03,GaliautdinovPRA07,GaliautdinovJMP07,GaliautdinovPRA09,PhysRevA.81.012320,geller2007springer}. In this work we analyze a CNOT gate based on conditional spectroscopy for a superconducting phase qubit coupled to a resonator. A related construction has already been demonstrated for a pair of superconducting flux qubits \cite{Mooij2007nature} and a similar concept has been explored in the context of NMR quantum computing \cite{Cory199882}. Although this approach can be applied to a variety of physical systems, the fidelity observed in the experiment  of Ref.~[\onlinecite{Mooij2007nature}] is not sufficient for practical use. Here we propose a method to improve the intrinsic fidelity and we then calculate the optimal fidelity as a function of total gate time. The phase qubit coupled to resonator system we consider is relevant to the UCSB Rezqu architecture. \footnote{J. M. Martinis, private communication.}

The idea behind a spectroscopic CNOT gate is simple and has a wide range of applicability: A $\pi$ pulse is applied
to the target qubit with a carefully selected carrier frequency $\omega_{\rm rf}$. The carrier frequency is close to the
qubit transition frequency given that the attached control resonator is in the ``on" or $|1\rangle$ state,  which in the 
$| qr\rangle$ basis is
\begin{equation}
\omega_{\rm on} \equiv \frac{E_{11}-E_{01}}{\hbar}.
\end{equation}
The Rabi frequency has to be smaller than the detuning to the ``off" transition at
\begin{equation}
\omega_{\rm off} \equiv \frac{E_{10}-E_{00}}{\hbar}.
\end{equation}
A direct $\sigma^z \otimes \sigma^z$ coupling between the devices would of course
generate a difference in $\omega_{\rm on}$ and $\omega_{\rm off}$, but in the phase qubit 
plus resonator system---which has no direct $\sigma^z \otimes \sigma^z$ coupling---such an
interaction is generated by level repulsion from the noncomputational $|2\rangle$
states. The difference $\big| \omega_{\rm on} - \omega_{\rm off}  \big|$ characterizes the 
sensitivity of the conditioning effect and determines the speed of the resulting gate. 

When the qubit and resonator are detuned by an amount larger than the coupling
between them, they become weakly coupled. In this limit, the sensitivity for current 
devices is limited to a few MHz, which is not sufficient for practical application. Therefore
to amplify the sensitivity we adiabatically bring the target qubit to a suitable point near resonance with the control resonator, and drive the qubit while it is strongly coupled with the resonator.
After performing a $\pi$ pulse the qubit is adiabatically detuned from the resonator.
 
Two main sources of intrinsic errors exist in this approach: Although we set the carrier frequency $\omega_{\rm rf}$ to a value such that we have a $\pi$ pulse in the qubit when the resonator is in $\ket{1}$, there is a small probability for the qubit to get rotated even if the resonator is in $\ket{0}$. The second error comes from the fact that, since we are driving the qubit, it is  possible to have leakage to the qubit  $\ket{2}$ state. The fidelity will reach its maximum value when both of these errors are minimized simultaneously. We use the DRAG method \cite{motzoi2009prl} to suppress the error due to leakage and adjust all other parameters by optimization. 
  
\section{CNOT Design}
\label{sec:gate}

\subsection{Hamiltonian}
In the basis of uncoupled qubits, the Hamiltonian of a qubit capacitively coupled to a resonator (assuming harmonic eigenfunction of 2-states) is given by (suppressing $\hbar$),
\BEq\label{hamiltonian}
H=H_{0}+H_{\rm rf}+H_{\rm int},
\EEq
where,
\BEq\label{hamiltonianParameters}
\begin{array}{l}
H_{0}=\left(\begin{array}{ccc}
0 & 0 & 0 \\
0 & \epsilon & 0 \\
0 & 0 & 2\epsilon-\Delta\end{array}\right)_{q}  + \left(\begin{array}{ccc}
0 & 0 & 0 \\
0 & \omega & 0 \\
0 & 0 & 2\omega\end{array}\right)_{r}\\
\\
H_{\rm rf}=\Omega(t)\cos(\omega_{\rm rf}(t)~t+\phi(t))\left(\begin{array}{ccc}
0 & 1 & 0 \\
1 & 0 & \sqrt{2} \\
0 & \sqrt{2} & 0\end{array}\right)_{q}\\
\\
H_{\rm int}=g \left(\begin{array}{ccc}
0 & -i & 0 \\
i & 0 & -\sqrt{2}i \\
0 & \sqrt{2}i & 0\end{array}\right)_{q}\otimes \left(\begin{array}{ccc}
0 & -i & 0 \\
i & 0 & -\sqrt{2}i \\
0 & \sqrt{2}i & 0\end{array}\right)_{r},
\end{array}
\EEq
where, $\epsilon$, $\omega$ and $\Delta$ are qubit frequency, resonator frequency, and anharmonicity of the qubit, respectively. $\Omega$, $\omega_{\rm rf}$ and $\phi$ are Rabi frequency, carrier frequency, and phase of the microwave pulse. $g$ is the (time-independent) interaction strength between qubit and resonator.

\begin{figure}[htb]
	\centering
\includegraphics[angle=0,width=1.00\linewidth]{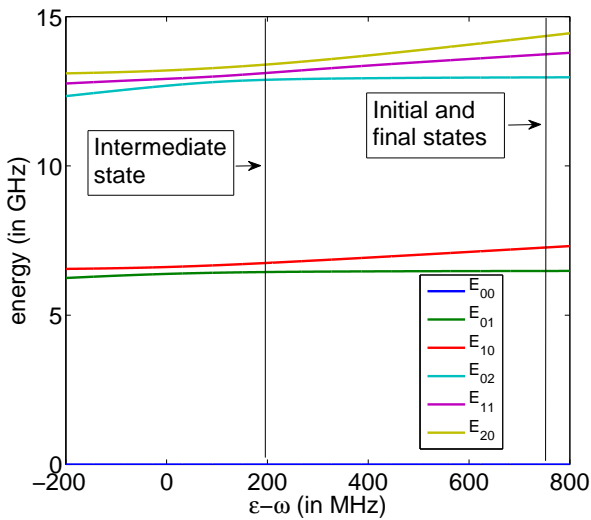}
				\caption{(Color online) A plot of first six energy levels of a phase qubit coupled to a resonator at $g/h=115$ MHz.}
		\label{fig:energy}
\end{figure}

The first six energy levels of the system (obtained numerically) are shown in Fig.(\ref{fig:energy}). We use coupling $g/h$=115 MHz, resonator frequency $\omega/h=6.5$ GHz and anharmonicity of the qubit $\Delta/h=200$ MHz. The simulations discussed below are carried out in a frame rotating with the instantaneous frequency of the qubit.

\subsection{CNOT Protocol}

Eigenstates of the full Hamiltonian reduce to the eigenstates of the uncoupled Hamiltonian far away from the resonance. We denote the first six eigenstates of the full Hamiltonian by {$E_{00}, E_{01}, E_{10}, E_{02}, E_{11}, E_{20}$} and define a conditional control sensitivity $S_{c}$ and leakage sensitivity $S_{l}$ as
\BEq\label{concoff}
\begin{array}{l}
S_{c}\equiv|(E_{11}-E_{01})-(E_{10}-E_{00})|, \\
S_{l}\equiv|(E_{21}-E_{11})-(E_{11}-E_{01})|.
\end{array}
\EEq
The conditional control sensitivity is the (magnitude of the) difference between $\omega_{\rm on}$ and $ \omega_{\rm off}$ . Leakage sensitivity is the anharmonicity of the target qubit when control resonator is on. In order to achieve a high  
fidelity both of these quantities need to be maximized (by varying the detuning $\epsilon-\omega$).

\begin{figure}[htb]
	\centering
\includegraphics[angle=0,width=1.00\linewidth]{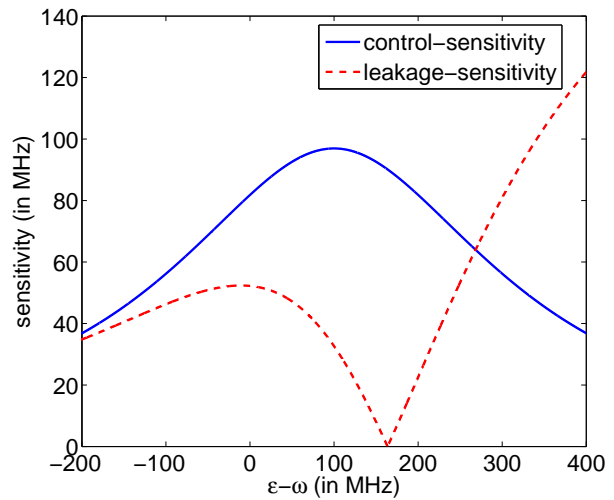}
				\caption{(Color online) A plot of control and leakage sensitivity vs. qubit frequency for $g/h=115$ MHz.}
		\label{fig:sensitivity}
\end{figure}

\begin{figure}[htb]
	\centering
\includegraphics[angle=0,width=1.00\linewidth]{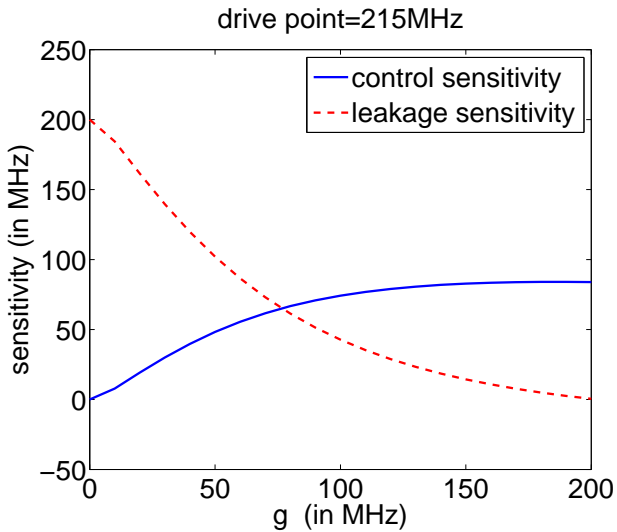}
				\caption{(Color online) A plot of control and leakage sensitivity vs. coupling.}
		\label{fig:coupling}
\end{figure}

A plot of these sensitivities is shown in Fig.~\ref{fig:sensitivity}. Peaks in the control sensitivity
(resulting from expected anticrossings) at detuning equal to zero and $\Delta$ conspire to give 
the maximum at 100 MHz detuning shown in the blue curve of Fig.~\ref{fig:sensitivity}.
Operating near 100 MHz detuning, however, leads to poor performance because of the large 
leakage error there. Better operation points exist near -100 and 200 MHz detuning; we shall 
make use of the latter. Fig.(\ref{fig:coupling}) shows the behavior of sensitivities vs. coupling at 
$(\epsilon-\omega)/h=215$ MHz. 

To implement a CNOT gate we begin with a strongly detuned qubit-resonator system.
Then the qubit is adiabatically tuned into resonance with the resonator, driven with a $\pi$ pulse, and
finally detuned. In the $\ket{qr}$ basis this protocol ideally produces
\BEq\label{ugate}
U_{\rm target}=\left[\begin{array}{cccc}
1 & 0 & 0 & 0 \\
0 & 0 & 0 & 1 \\
0 & 0 & 1 & 0 \\
0 & 1 & 0 & 0 \end{array}\right]\\
= {\rm SWAP \times CNOT \times SWAP},
\EEq
where
\begin{equation}
{\rm SWAP} \equiv
\left[\begin{array}{cccc}
1 & 0 & 0 & 0 \\
0 & 0 & 1 & 0 \\
0 & 1 & 0 & 0 \\
0 & 0 & 0 & 1 \end{array}\right]\\
\end{equation}
 is the swap gate. To obtain this target we also perform $z$ rotations on the qubit
before and after the sequence described above, with angles determined by optimizations.

\subsection{Local Rotations with DRAG}

As far as leakage outside the computational subspace is concerned, we can consider $E_{01}$,$E_{11}$ and $E_{21}$ to be a single 3-level quantum system where we are interested in local rotations between the first two levels and therefore, in order to suppress the errors due to leakage to the third level, local rotations are performed with a DRAG pulse \cite{motzoi2009prl}, up to $5^{\rm th}$ order. In order to do a local rotation of angle $\theta$ about the $x$ axis, we set the Rabi pulse of the Hamiltonian to be

\BEq\label{drag1}
\begin{array}{l}
\Omega(t)\cos(\phi(t))=f_{\rm \theta}+\frac{(\lambda^{2}-4)f_{\rm \theta}^3}{8{\Delta}^2}-\frac{(13\lambda^4-76\lambda^2+112)f_{\rm \theta}^5}{128{\Delta}^4},\\
\\
\Omega(t)\sin(\phi(t))=-\frac{\dot{f_{\rm \theta}}}{\Delta}+\frac{33(\lambda^2-2)f_{\rm \theta}^{2}\dot{f_{\rm \theta}}}{24\Delta^3},\\
\\
\omega_{\rm rf}(t)=\omega_{c}+\frac{(\lambda^2-4)f_{\rm \theta}^2}{4\Delta}-\frac{(\lambda^4-7\lambda^2+12)f_{\rm \theta}^4}{16{\Delta}^3},
\end{array}
\EEq
where, $\lambda=\sqrt{2}$ for our case and the first two equations give the amplitudes of $\cos(\omega_{\rm rf} t)$ and $\sin(\omega_{\rm rf} t)$ quadratures. Here $\omega_{c}$ is found via optimization around $|E_{01}$-$E_{11}|$ and $f_{\rm \theta}(t)$ is chosen to be a Gaussian function (vertically shifted) such that $f_{\rm \theta}(0)=f_{\rm \theta}(t_{\rm g})=0$ and $\displaystyle\int\limits_{0}^{t_{\rm g}}f_{\rm \theta}(t)dt=\theta$, $t_{\rm g}$ being the time required to perform the local rotation.

\section{Gate Optimization}
\label{sec:sim}

In this section we describe the sources of error and our methodology to compute the intrinsic  fidelity curve.
As mentioned above, there are two major intrinsic sources of error.
In our simulation we use the full nine-level Hamiltonian to compute the fidelity curve and use DRAG to suppress the leakage error while the control error is treated with an adjustment of other controllable parameters via optimization.

The state-dependent process fidelity between two quantum gates $U$ and $U_{\rm target}$ is given by,
\begin{equation}
F(U_{\rm target},U,\ket{\psi})\equiv\big| \langle \psi | U_{\rm target}^\dagger U | \psi \rangle \big|^2.
\end{equation}
The fidelity measure we use in this work is the process fidelity averaged over the four dimensional Hilbert space, that can be shown \cite{Pedersen20087028,PhysRevA.81.052340} to be exactly equal to a trace formula given by,  
\begin{equation} \label{f}
F\left(U_{\rm target},U\right) \equiv \frac{{\rm Tr}(UU^{\dagger})+\left|{\rm Tr} \, (U_{\rm target}^{\dagger}U)\right|^{2}}{20},
\end{equation}
where $U_{\rm target}$ is always four dimensional for a two-qubit operation and $U$ is the time evolution operator projected into the four dimensional computational subspace of the entire Hilbert space. This definition 
assumes that $U_{\rm target}$ is unitary, but $U$ does not have to be (because the final evolution operator after projection
into the four-dimensional computational subspace is not necessarily unitary).

\begin{figure}[htb]
	\centering
\includegraphics[angle=0,width=1.00\linewidth]{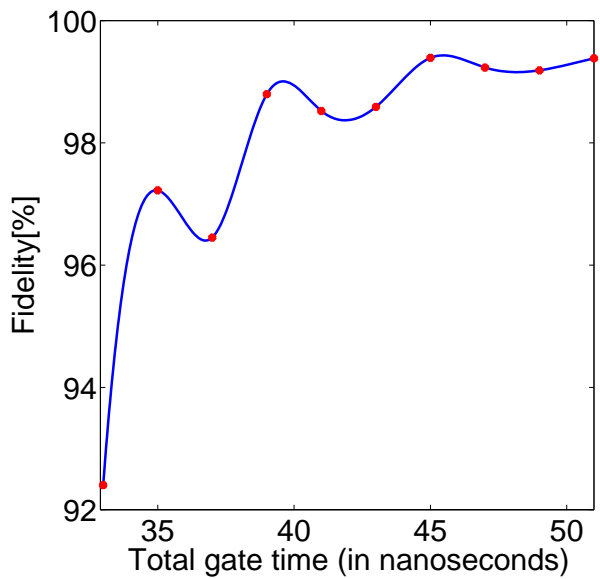}
				\caption{(Color online) A plot of fidelity vs. total gate time in nanoseconds. Red points are obtained via optimization and the blue line is an interpolation.}
		\label{fig:fidcurve}
\end{figure}

In order to compute the fidelity curve, we perform a 8 dimensional optimization over the control parameters for a given total gate time and obtain the best fidelity. The 8 control parameters optimized here are coupling ($g$), qubit frequency in the intermediate state (see Fig.\ref{fig:energy}), carrier frequency of the Rabi pulse ($\omega_{c}$), angle of rotation by the Rabi pulse, duration of each ramp pulse ($t_{\rm ramp}$) and three other parameters related to the areas of the gaussian envelope of DRAG pulse. The result obtained from such an optimization corresponds to a single point in the plot of Fig.(\ref{fig:fidcurve}).  In order to help to control the phases of the matrix elements developed in the time-evolution operator, we also attach a pre and a post z-rotation of the qubit. We observe from Fig.(\ref{fig:sensitivity}) and Fig.(\ref{fig:coupling}) that a good range for driving point ($\epsilon-\omega$) should be $\thicksim$ 200-250 MHz and a good range for coupling strength should be $\thicksim$ 100-125 MHz. These numbers are used as guess values of our multidimensional optimization search. 

Fig.(\ref{fig:fidcurve}) shows the fidelity curve obtained from optimization and Table(\ref{paramtable}) shows corresponding coupling strengths, duration of each ramp and driving points. Our result shows that the intrinsic fidelity can be pushed to 99\% within 45 ns gate time. The remaining error comes from the adiabaticity of the ramp pulses and leakage outside the computational basis states, for example between $\ket{11}$ and $\ket{20}$ (in $\ket{\rm qr}$ basis) at driving point where $\epsilon \approx \omega+\Delta$.

\begin{table}[tb]
\centering
\caption{\label{paramtable}Optimum values of parameters for CNOT gate.}
\begin{tabular}{||c | c | c | c | c ||}
\hline & \ & \ & \ & \\
$t_{\rm gate}$ (ns.) & $t_{\rm ramp}$ (ns.) & $g$ (MHz) & $\epsilon-\omega$ (MHz) & Fidelity[\%]\\
\hline & \ & \ & \ & \\
   33 & 7.1 & 118.5966 & 235.1804 & 92.4021 \\
   35 & 7.3 & 107.5055 & 218.8692 & 97.2223 \\
   37 & 7.0 & 123.8367 & 237.4236 & 96.4515 \\
   39 & 7.3 & 113.2576 & 217.7214 & 98.7965 \\
   41 & 7.4 & 107.7414 & 218.2388 & 98.5218 \\ 
   43 & 7.2 & 117.6687 & 218.6669 & 98.5886 \\
   45 & 7.5 & 107.3625 & 199.2662 & 99.3914 \\
   47 & 7.5 & 104.0402 & 204.6788 & 99.2326 \\
   49 & 7.3 & 113.8836 & 208.3964 & 99.1864 \\
   51 & 7.5 & 107.6466 & 202.2407 & 99.3836 \\
& \ & \ & \ & \\
\hline
\end{tabular}
\end{table}

\begin{figure}[htb]
	\centering
\includegraphics[angle=0,width=1.00\linewidth]{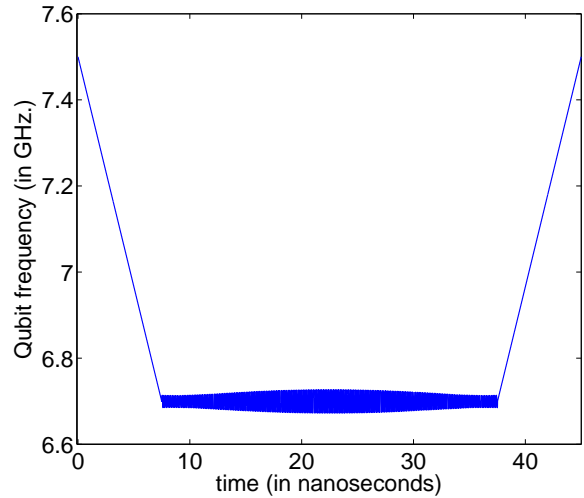}
				\caption{(Color online) A plot of qubit frequency vs. time for CNOT at total gate time=45 ns.}
		\label{fig:pulseshape}
\end{figure}

As an example, we show the change of qubit frequency (in GHz) over time (in nanoseconds) in Fig.(\ref{fig:pulseshape}) for the CNOT having total gate time = 45 ns. while the resonator frequency is always fixed at 6.5 GHz and pre and post z-rotation angles are found to be $\vartheta_{\rm post}=-1.0915$ and $\vartheta_{\rm pre}=0.5442$ radian for this case. We use linear pulse for ramps and a gaussian envelope is used for DRAG.

\section{Conclusions}
\label{sec:con}

We have computed intrinsic fidelity of a CNOT gate based on conditional spectroscopy approach and have shown that it is possible to achieve greater than 99\% fidelity within an experimentally practical time scale of 45ns.
However,  this design requires a large coupling strength, so tunable coupling \cite{PhysRevB.82.104522}  would probably be required in a multi-qubit system.
Although our analysis assumed a phase qubit and resonator, the design also apples to capacitively coupled phase qubits,
where a slightly higher fidelity would be expected because of the additional anharmonicity. 

\begin{acknowledgments}
This work was supported by IARPA under award W911NF-04-1-0204.
It is a pleasure to thank John Martinis and Emily Pritchett for useful discussions.
\end{acknowledgments}

\bibliography{spectroscopicCNOT}

\begin{thebibliography}{17}%
\makeatletter
\providecommand \@ifxundefined [1]{%
 \@ifx{#1\undefined}
}%
\providecommand \@ifnum [1]{%
 \ifnum #1\expandafter \@firstoftwo
 \else \expandafter \@secondoftwo
 \fi
}%
\providecommand \@ifx [1]{%
 \ifx #1\expandafter \@firstoftwo
 \else \expandafter \@secondoftwo
 \fi
}%
\providecommand \natexlab [1]{#1}%
\providecommand \enquote  [1]{``#1''}%
\providecommand \bibnamefont  [1]{#1}%
\providecommand \bibfnamefont [1]{#1}%
\providecommand \citenamefont [1]{#1}%
\providecommand \href@noop [0]{\@secondoftwo}%
\providecommand \href [0]{\begingroup \@sanitize@url \@href}%
\providecommand \@href[1]{\@@startlink{#1}\@@href}%
\providecommand \@@href[1]{\endgroup#1\@@endlink}%
\providecommand \@sanitize@url [0]{\catcode `\\12\catcode `\$12\catcode
  `\&12\catcode `\#12\catcode `\^12\catcode `\_12\catcode `\%12\relax}%
\providecommand \@@startlink[1]{}%
\providecommand \@@endlink[0]{}%
\providecommand \url  [0]{\begingroup\@sanitize@url \@url }%
\providecommand \@url [1]{\endgroup\@href {#1}{\urlprefix }}%
\providecommand \urlprefix  [0]{URL }%
\providecommand \Eprint [0]{\href }%
\@ifxundefined \urlstyle {%
  \providecommand \doi  [0]{\begingroup \@sanitize@url \@doi}%
  \providecommand \@doi [1]{\endgroup \@@startlink {\doibase
  #1}doi:\discretionary {}{}{}#1\@@endlink }%
}{%
  \providecommand \doi  [0]{doi:\discretionary{}{}{}\begingroup
  \urlstyle{rm}\Url }%
}%
\providecommand \doibase [0]{http://dx.doi.org/}%
\providecommand \Doi [0]{\begingroup \@sanitize@url \@Doi }%
\providecommand \@Doi  [1]{\endgroup\@@startlink{\doibase#1}\@@Doi}%
\providecommand \@@Doi [1]{#1\@@endlink}%
\providecommand \selectlanguage [0]{\@gobble}%
\providecommand \bibinfo  [0]{\@secondoftwo}%
\providecommand \bibfield  [0]{\@secondoftwo}%
\providecommand \translation [1]{[#1]}%
\providecommand \BibitemOpen [0]{}%
\providecommand \bibitemStop [0]{}%
\providecommand \bibitemNoStop [0]{.\EOS\space}%
\providecommand \EOS [0]{\spacefactor3000\relax}%
\providecommand \BibitemShut  [1]{\csname bibitem#1\endcsname}%
\bibitem [{\citenamefont {Nielsen}\ and\ \citenamefont
  {Chuang}(2005)}]{ncbook}%
  \BibitemOpen
  \bibfield  {author} {\bibinfo {author} {\bibfnamefont {M.~A.}\ \bibnamefont
  {Nielsen}}\ and\ \bibinfo {author} {\bibfnamefont {I.~L.}\ \bibnamefont
  {Chuang}},\ }\href@noop {} {\emph {\bibinfo {title} {Quantum Computation and
  Quantum Information}}}\ (\bibinfo  {publisher} {Cambridge University Press},\
  \bibinfo {year} {2005})\BibitemShut {NoStop}%
\bibitem [{\citenamefont {DiVincenzo}(2000)}]{DiVincenzo2000fp}%
  \BibitemOpen
  \bibfield  {author} {\bibinfo {author} {\bibfnamefont {David~P.}\
  \bibnamefont {DiVincenzo}},\ }\bibfield  {title} {\enquote {\bibinfo {title}
  {The physical implementation of quantum computation},}\ }\href@noop {}
  {\bibfield  {journal} {\bibinfo  {journal} {Fortschr. Phys.},\ }\textbf
  {\bibinfo {volume} {48}},\ \bibinfo {pages} {771} (\bibinfo {year}
  {2000})}\BibitemShut {NoStop}%
\bibitem [{\citenamefont {You}\ and\ \citenamefont {Nori}(2005)}]{You2005pt}%
  \BibitemOpen
  \bibfield  {author} {\bibinfo {author} {\bibfnamefont {J.~Q.}\ \bibnamefont
  {You}}\ and\ \bibinfo {author} {\bibfnamefont {F.}~\bibnamefont {Nori}},\
  }\bibfield  {title} {\enquote {\bibinfo {title} {Superconducting circuits and
  quantum information},}\ }\href@noop {} {\bibfield  {journal} {\bibinfo
  {journal} {Physics Today},\ }\textbf {\bibinfo {volume} {58}},\ \bibinfo
  {pages} {42} (\bibinfo {year} {2005})},\ \bibinfo {note} {p. 42}\BibitemShut
  {NoStop}%
\bibitem [{\citenamefont {Strauch}\ \emph {et~al.}(2003)\citenamefont
  {Strauch}, \citenamefont {Johnson}, \citenamefont {Dragt}, \citenamefont
  {Lobb}, \citenamefont {Anderson},\ and\ \citenamefont
  {Wellstood}}]{StrauchPRL03}%
  \BibitemOpen
  \bibfield  {author} {\bibinfo {author} {\bibfnamefont {F.~W.}\ \bibnamefont
  {Strauch}}, \bibinfo {author} {\bibfnamefont {P.~R.}\ \bibnamefont
  {Johnson}}, \bibinfo {author} {\bibfnamefont {A.~J.}\ \bibnamefont {Dragt}},
  \bibinfo {author} {\bibfnamefont {C.~J.}\ \bibnamefont {Lobb}}, \bibinfo
  {author} {\bibfnamefont {J.~R.}\ \bibnamefont {Anderson}}, \ and\ \bibinfo
  {author} {\bibfnamefont {F.~C.}\ \bibnamefont {Wellstood}},\ }\bibfield
  {title} {\enquote {\bibinfo {title} {Quantum logic gates for coupled
  superconducting phase qubits},}\ }\href@noop {} {\bibfield  {journal}
  {\bibinfo  {journal} {Phys. Rev. Lett.},\ }\textbf {\bibinfo {volume} {91}},\
  \bibinfo {pages} {167005} (\bibinfo {year} {2003})}\BibitemShut {NoStop}%
\bibitem [{\citenamefont {Zhang}\ \emph {et~al.}(2003)\citenamefont {Zhang},
  \citenamefont {Vala}, \citenamefont {Sastry},\ and\ \citenamefont
  {Whaley}}]{ZhangPRA03}%
  \BibitemOpen
  \bibfield  {author} {\bibinfo {author} {\bibfnamefont {J.}~\bibnamefont
  {Zhang}}, \bibinfo {author} {\bibfnamefont {J.}~\bibnamefont {Vala}},
  \bibinfo {author} {\bibfnamefont {S.}~\bibnamefont {Sastry}}, \ and\ \bibinfo
  {author} {\bibfnamefont {K.~B.}\ \bibnamefont {Whaley}},\ }\bibfield  {title}
  {\enquote {\bibinfo {title} {Geometric theory of nonlocal two-qubit
  operations},}\ }\href@noop {} {\bibfield  {journal} {\bibinfo  {journal}
  {Phys. Rev. A},\ }\textbf {\bibinfo {volume} {67}},\ \bibinfo {pages}
  {042313} (\bibinfo {year} {2003})}\BibitemShut {NoStop}%
\bibitem [{\citenamefont
  {Galiautdinov}(2007){\natexlab{a}}}]{GaliautdinovPRA07}%
  \BibitemOpen
  \bibfield  {author} {\bibinfo {author} {\bibfnamefont {A.}~\bibnamefont
  {Galiautdinov}},\ }\bibfield  {title} {\enquote {\bibinfo {title} {Generation
  of high-fidelity controlled-{NOT} logic gates by coupled superconducting
  qubits},}\ }\href@noop {} {\bibfield  {journal} {\bibinfo  {journal} {Phys.
  Rev. A},\ }\textbf {\bibinfo {volume} {75}},\ \bibinfo {pages} {052303}
  (\bibinfo {year} {2007}{\natexlab{a}})}\BibitemShut {NoStop}%
\bibitem [{\citenamefont
  {Galiautdinov}(2007){\natexlab{b}}}]{GaliautdinovJMP07}%
  \BibitemOpen
  \bibfield  {author} {\bibinfo {author} {\bibfnamefont {A.}~\bibnamefont
  {Galiautdinov}},\ }\bibfield  {title} {\enquote {\bibinfo {title}
  {Single-step controlled-{NOT} logic from any exchange interaction},}\
  }\href@noop {} {\bibfield  {journal} {\bibinfo  {journal} {J. Math. Phys.},\
  }\textbf {\bibinfo {volume} {48}},\ \bibinfo {pages} {112105} (\bibinfo
  {year} {2007}{\natexlab{b}})}\BibitemShut {NoStop}%
\bibitem [{\citenamefont {Galiautdinov}(2009)}]{GaliautdinovPRA09}%
  \BibitemOpen
  \bibfield  {author} {\bibinfo {author} {\bibfnamefont {A.}~\bibnamefont
  {Galiautdinov}},\ }\bibfield  {title} {\enquote {\bibinfo {title}
  {Controlled-{NOT} logic with nonresonant {J}osephson phase qubits},}\
  }\href@noop {} {\bibfield  {journal} {\bibinfo  {journal} {Phys. Rev. A},\
  }\textbf {\bibinfo {volume} {79}},\ \bibinfo {pages} {042316} (\bibinfo
  {year} {2009})}\BibitemShut {NoStop}%
\bibitem [{\citenamefont {Geller}\ \emph {et~al.}(2010)\citenamefont {Geller},
  \citenamefont {Pritchett}, \citenamefont {Galiautdinov},\ and\ \citenamefont
  {Martinis}}]{PhysRevA.81.012320}%
  \BibitemOpen
  \bibfield  {author} {\bibinfo {author} {\bibfnamefont {Michael~R.}\
  \bibnamefont {Geller}}, \bibinfo {author} {\bibfnamefont {Emily~J.}\
  \bibnamefont {Pritchett}}, \bibinfo {author} {\bibfnamefont {Andrei}\
  \bibnamefont {Galiautdinov}}, \ and\ \bibinfo {author} {\bibfnamefont
  {John~M.}\ \bibnamefont {Martinis}},\ }\bibfield  {title} {\enquote {\bibinfo
  {title} {Quantum logic with weakly coupled qubits},}\ }\Doi
  {10.1103/PhysRevA.81.012320} {\bibfield  {journal} {\bibinfo  {journal}
  {Phys. Rev. A},\ }\textbf {\bibinfo {volume} {81}},\ \bibinfo {pages}
  {012320} (\bibinfo {year} {2010})}\BibitemShut {NoStop}%
\bibitem [{\citenamefont {Geller}\ \emph {et~al.}(2007)\citenamefont {Geller},
  \citenamefont {Pritchett}, \citenamefont {Sornborger},\ and\ \citenamefont
  {Wilhelm}}]{geller2007springer}%
  \BibitemOpen
  \bibfield  {author} {\bibinfo {author} {\bibfnamefont {M.}~\bibnamefont
  {Geller}}, \bibinfo {author} {\bibfnamefont {E.}~\bibnamefont {Pritchett}},
  \bibinfo {author} {\bibfnamefont {A.}~\bibnamefont {Sornborger}}, \ and\
  \bibinfo {author} {\bibfnamefont {F.}~\bibnamefont {Wilhelm}},\ }\bibfield
  {title} {\enquote {\bibinfo {title} {Quantum computing with superconductors
  i: Architectures},}\ }in\ \href@noop {} {\emph {\bibinfo {booktitle}
  {Manipulating Quantum Coherence in Solid State Systems}}},\ \bibinfo {series}
  {NATO Science Series}, Vol.\ \bibinfo {volume} {244},\ \bibinfo {editor}
  {edited by\ \bibinfo {editor} {\bibfnamefont {Michael}\ \bibnamefont
  {Flatte}}\ and\ \bibinfo {editor} {\bibfnamefont {I.}~\bibnamefont
  {Tifrea}}}\ (\bibinfo  {publisher} {Springer Netherlands},\ \bibinfo {year}
  {2007})\ pp.\ \bibinfo {pages} {171--194}\BibitemShut {NoStop}%
\bibitem [{\citenamefont {Plantenberg}\ \emph {et~al.}(2007)\citenamefont
  {Plantenberg}, \citenamefont {de~Groot}, \citenamefont {Harmans},\ and\
  \citenamefont {Mooij}}]{Mooij2007nature}%
  \BibitemOpen
  \bibfield  {author} {\bibinfo {author} {\bibfnamefont {J.~H.}\ \bibnamefont
  {Plantenberg}}, \bibinfo {author} {\bibfnamefont {P.~C.}\ \bibnamefont
  {de~Groot}}, \bibinfo {author} {\bibfnamefont {C.~J. P.~M}\ \bibnamefont
  {Harmans}}, \ and\ \bibinfo {author} {\bibfnamefont {J.~E.}\ \bibnamefont
  {Mooij}},\ }\bibfield  {title} {\enquote {\bibinfo {title} {Demonstration of
  controlled-not quantum gates on a pair of superconducting quantum bits},}\
  }\href@noop {} {\bibfield  {journal} {\bibinfo  {journal} {Nature},\ }\textbf
  {\bibinfo {volume} {447}},\ \bibinfo {pages} {836} (\bibinfo {year}
  {2007})}\BibitemShut {NoStop}%
\bibitem [{\citenamefont {Cory}\ \emph {et~al.}(1998)\citenamefont {Cory},
  \citenamefont {Price},\ and\ \citenamefont {Havel}}]{Cory199882}%
  \BibitemOpen
  \bibfield  {author} {\bibinfo {author} {\bibfnamefont {David~G.}\
  \bibnamefont {Cory}}, \bibinfo {author} {\bibfnamefont {Mark~D.}\
  \bibnamefont {Price}}, \ and\ \bibinfo {author} {\bibfnamefont {Timothy~F.}\
  \bibnamefont {Havel}},\ }\bibfield  {title} {\enquote {\bibinfo {title}
  {Nuclear magnetic resonance spectroscopy: An experimentally accessible
  paradigm for quantum computing},}\ }\Doi {DOI: 10.1016/S0167-2789(98)00046-3}
  {\bibfield  {journal} {\bibinfo  {journal} {Physica D: Nonlinear Phenomena},\
  }\textbf {\bibinfo {volume} {120}},\ \bibinfo {pages} {82 -- 101} (\bibinfo
  {year} {1998})},\ ISSN \bibinfo {issn} {0167-2789},\ \bibinfo {note}
  {proceedings of the Fourth Workshop on Physics and Consumption}\BibitemShut
  {NoStop}%
\bibitem [{Note1()}]{Note1}%
  \BibitemOpen
  \bibinfo {note} {J. M. Martinis, private communication.}\BibitemShut {Stop}%
\bibitem [{\citenamefont {Motzoi}\ \emph {et~al.}(2009)\citenamefont {Motzoi},
  \citenamefont {Gambetta}, \citenamefont {Rebentrost},\ and\ \citenamefont
  {Wilhelm}}]{motzoi2009prl}%
  \BibitemOpen
  \bibfield  {author} {\bibinfo {author} {\bibfnamefont {F.}~\bibnamefont
  {Motzoi}}, \bibinfo {author} {\bibfnamefont {J.~M.}\ \bibnamefont
  {Gambetta}}, \bibinfo {author} {\bibfnamefont {P.}~\bibnamefont
  {Rebentrost}}, \ and\ \bibinfo {author} {\bibfnamefont {F.~K.}\ \bibnamefont
  {Wilhelm}},\ }\bibfield  {title} {\enquote {\bibinfo {title} {Simple pulses
  for elimination of leakage in weakly nonlinear qubits},}\ }\Doi
  {10.1103/PhysRevLett.103.110501} {\bibfield  {journal} {\bibinfo  {journal}
  {Phys. Rev. Lett.},\ }\textbf {\bibinfo {volume} {103}},\ \bibinfo {pages}
  {110501} (\bibinfo {year} {2009})}\BibitemShut {NoStop}%
\bibitem [{\citenamefont {Pedersen}\ \emph {et~al.}(2008)\citenamefont
  {Pedersen}, \citenamefont {M{\o}ller},\ and\ \citenamefont
  {M{\o}lmer}}]{Pedersen20087028}%
  \BibitemOpen
  \bibfield  {author} {\bibinfo {author} {\bibfnamefont {Line~Hjortsh{\o}j}\
  \bibnamefont {Pedersen}}, \bibinfo {author} {\bibfnamefont {Niels~Martin}\
  \bibnamefont {M{\o}ller}}, \ and\ \bibinfo {author} {\bibfnamefont {Klaus}\
  \bibnamefont {M{\o}lmer}},\ }\bibfield  {title} {\enquote {\bibinfo {title}
  {The distribution of quantum fidelities},}\ }\Doi {DOI:
  10.1016/j.physleta.2008.10.034} {\bibfield  {journal} {\bibinfo  {journal}
  {Physics Letters A},\ }\textbf {\bibinfo {volume} {372}},\ \bibinfo {pages}
  {7028 -- 7032} (\bibinfo {year} {2008})},\ ISSN \bibinfo {issn}
  {0375-9601}\BibitemShut {NoStop}%
\bibitem [{\citenamefont {Ghosh}\ and\ \citenamefont
  {Geller}(2010)}]{PhysRevA.81.052340}%
  \BibitemOpen
  \bibfield  {author} {\bibinfo {author} {\bibfnamefont {Joydip}\ \bibnamefont
  {Ghosh}}\ and\ \bibinfo {author} {\bibfnamefont {Michael~R.}\ \bibnamefont
  {Geller}},\ }\bibfield  {title} {\enquote {\bibinfo {title} {Controlled-not
  gate with weakly coupled qubits: Dependence of fidelity on the form of
  interaction},}\ }\Doi {10.1103/PhysRevA.81.052340} {\bibfield  {journal}
  {\bibinfo  {journal} {Phys. Rev. A},\ }\textbf {\bibinfo {volume} {81}},\
  \bibinfo {pages} {052340} (\bibinfo {year} {2010})}\BibitemShut {NoStop}%
\bibitem [{\citenamefont {Pinto}\ \emph {et~al.}(2010)\citenamefont {Pinto},
  \citenamefont {Korotkov}, \citenamefont {Geller}, \citenamefont {Shumeiko},\
  and\ \citenamefont {Martinis}}]{PhysRevB.82.104522}%
  \BibitemOpen
  \bibfield  {author} {\bibinfo {author} {\bibfnamefont {Ricardo~A.}\
  \bibnamefont {Pinto}}, \bibinfo {author} {\bibfnamefont {Alexander~N.}\
  \bibnamefont {Korotkov}}, \bibinfo {author} {\bibfnamefont {Michael~R.}\
  \bibnamefont {Geller}}, \bibinfo {author} {\bibfnamefont {Vitaly~S.}\
  \bibnamefont {Shumeiko}}, \ and\ \bibinfo {author} {\bibfnamefont {John~M.}\
  \bibnamefont {Martinis}},\ }\bibfield  {title} {\enquote {\bibinfo {title}
  {Analysis of a tunable coupler for superconducting phase qubits},}\ }\Doi
  {10.1103/PhysRevB.82.104522} {\bibfield  {journal} {\bibinfo  {journal}
  {Phys. Rev. B},\ }\textbf {\bibinfo {volume} {82}},\ \bibinfo {pages}
  {104522} (\bibinfo {year} {2010})}\BibitemShut {NoStop}%
\end{thebibliography}%

\end{document}